\documentclass[aip, jcp, reprint, nolinenumbers, twocolumn, nobalancelastpage, nofootinbib]{revtex4-1}

\usepackage{todonotes}
\usepackage{soul}
\usepackage{enumitem}
\usepackage{graphicx}          
\usepackage{dcolumn}           
\usepackage{bm}                
\usepackage[mathlines]{lineno} 
\usepackage[utf8]{inputenc}
\usepackage{mathrsfs}
\usepackage[version=4]{mhchem}
\usepackage[]{algorithm}
\usepackage{algpseudocode}
\usepackage{amsmath}    
\usepackage{amssymb}    
\usepackage{amsthm}
\usepackage{graphicx}   
\usepackage{verbatim}   
\usepackage{color}      
\usepackage{subfigure}  
\usepackage[capitalise]{cleveref}   
\usepackage{appendix}

\usepackage{alphalph}
\makeatletter
\newcommand*{\fnsymbolsingle}[1]{%
  \ensuremath{%
    \ifcase#1%
    \or *%
    \or \dagger
    \or \ddagger
    \or \mathsection
    \or \mathparagraph
    \else
      \@ctrerr
    \fi
  }%
}
\makeatother
\newalphalph{\fnsymbolmult}[mult]{\fnsymbolsingle}{}

\newcommand{\beginsupplement}{%
        \setcounter{table}{0}
        \renewcommand{\thetable}{S\arabic{table}}%
        \setcounter{figure}{0}
        \renewcommand{\thefigure}{S\arabic{figure}}%
     }

\begin{document}

\title{A maximum-caliber approach to predicting perturbed folding kinetics due to mutations}

\author{Vincent A. Voelz}
\email{vvoelz@temple.edu}
\author{Guangfeng Zhou}
\author{Hongbin Wan}
\affiliation{Department of Chemistry, Temple University, Philadelphia, PA 19122, USA}

\begin{abstract}
We present a maximum-caliber method for inferring transition rates of a Markov State Model (MSM) with perturbed equilibrium populations, given estimates of state populations and rates for an unperturbed MSM.  It is similar in spirit to previous approaches but given the inclusion of prior information it is more robust and simple to implement.  We examine its performance in simple biased diffusion models of kinetics, and then apply the method to predicting changes in folding rates for several highly non-trivial protein folding systems for which non-native interactions play a significant role, including (1) tryptophan variants of GB1 hairpin, (2) salt-bridge mutations of Fs peptide helix, and (3) MSMs built from ultra-long folding trajectories of FiP35 and GTT variants of WW domain.  In all cases, the method correctly predicts changes in folding rates, suggesting the wide applicability of maximum-caliber approaches to efficiently predict how mutations perturb protein conformational dynamics. 
\end{abstract}
\maketitle





\section{Introduction}

Markov State Models (MSMs) of conformational dynamics have significantly advanced our understanding of biomolecular function.\cite{Kohlhoff:2013bo,Plattner:2015br,Voelz:2010hs} In the MSM approach, conformational dynamics is modeled as a kinetic network of transitions between metastable states.\cite{Chodera:2014gk,Prinz:2011id}  This analysis works particularly well with large collections of molecular simulation trajectories obtained from parallel distributed computing,\cite{Shirts:2000uc,Buch:2010iq} which enable the identification of relevant metastable states and efficient sampling of the transitions rates between them.\cite{Noe:2009en}  A complete description of conformational dynamics comes from the estimates of discrete-time transition probabilities $p_{ij}$, the probability of transitioning between states $i$ and $j$ in some time interval $\tau$. 

Recently, so-called maximum-caliber approaches\cite{Presse:2013dh} have been used to infer transition rates $p_{ij}$ given only the equilibrium populations $\pi_i$ and constraints based on dynamical averages across the set of transitions.\cite{Dixit:2015bn,Dixit:2014hz}  This method works by maximizing the path entropy
\begin{equation}
\mathcal{S} = - \sum_{i,j} \pi_i p_{ij} \ln p_{ij},
\end{equation}
of a Markov State Model in the presence of constraints to enforce the conservation of transition probabilities and dynamical averages. The result is obtained using a Lagrange multiplier approach.  

A natural application of the maximum-caliber method would be in trying to understand how the thermodynamic landscapes of proteins shape their kinetics.  For example, if we have an MSM of a given protein sequence, we would like to apply the maximum-caliber approach to a mutant protein sequence to infer differences in the folding kinetics directly from predicted changes in thermodynamics.  This is easily stated, but it is hard to achieve good results in practice. As described in Dixit et al.\cite{Dixit:2015bn} current maximum-caliber approaches require constraints on average dynamical quantities on which the results sensitively depend.   For processes like diffusion, a restraint on the mean jump rate leads to very good estimates of microscopic rates, but for systems like proteins, the microscopic rates between metastable states depends sensitively on highly local features, such as the kinetic barriers separating states and their local diffusivity.  Selection of meaningful dynamical restraints thus depends on obtaining a projection of the system allowing good estimation of the effective kinetic distance between metastable states.  In Dixit et al.,\cite{Dixit:2015bn}  simple geometric distance metrics fail to produce good maximum-caliber models from simulation data; instead, metrics that better capture effective kinetic distance prevail.\cite{Dixit:2015bn}   Similarly, the accuracy of an MSM depends strongly on our ability to find coordinate transformations which correctly capture the kinetic distance between states.\cite{Noe:2015km}. Fortunately, recent advances have enabled the development of methods that can capture kinetic distance very well, by exploiting the variational approach to conformational dynamics (VAC).\cite{Noe:2013kua,Nuske:2014kk,McGibbon:2015gm}.  These methods include time-lagged independent component analysis (tICA), which finds the linear combinations of structural order parameters that best capture long-timescale dynamics,\cite{Schwantes:2013bp,PerezHernandez:2013tt,Schwantes:2015if} and related Diffusion Map methods.\cite{Boninsegna:2015cz}

Instead of enforcing constraints on average dynamical quantities as in Dixit et al.,\cite{Dixit:2015bn} here we consider an unperturbed Markov State Model that has already been constructed for a given protein, and we next want to understand the kinetic behavior of a perturbed MSM, say, for a related sequence variant.  In this case, the existing MSM (if well-constructed) already tells us a great deal about the local kinetic environment shaping the rates between any two metastable states $i$ and $j$; we know the unperturbed equilibrium populations $\pi_i$ and $\pi_j$, as well as rate estimates $p_{ij}$ and $p_{ji}$, and hence how the thermodynamic gradient between states $i$ and $j$ is related to the rate of population flow between them.  It would be advantageous to use this information as a starting point for making rate estimates for similar systems.

In this paper, we present a maximum-caliber approach for inferring transition rates of a MSM with perturbed equilibrium populations that is robust and simple to implement.  We first show how the method performs in a simple biased diffusion system.   We then apply the method to predicting changes in folding rates for several highly non-trivial mini-protein folding systems in which non-native interactions play a significant role, including tryptophan mutants of GB1 hairpin, and salt-bridge mutations of Fs peptide. Finally, we apply the method to MSMs built from ultra-long folding trajectories of the GTT and Fip35 variants of WW domain from Shaw et al.\cite{Shaw:2010ge,LindorffLarsen:2011gl}   The the overall accuracy of these predictions suggest the wide applicability of maximum-caliber approach to predicting perturbed kinetics.

\section{Theory}

Suppose an MSM, defined by transition probabilities $p_{ij}^*$ and equilibrium populations $\pi_i^*$, is perturbed such that new equilibrium populations $\pi_i$ are known.  Our goal is to infer the new transition probabilities $p_{ij}$.  

To infer the values $p_{ij}$, we propose maximizing the path entropy, or caliber $\mathcal{C}$, of the perturbed MSM with respect to the unperturbed MSM, defined by 

\begin{equation}
\mathcal{C} = \sum_{i,j} - \pi_i p_{ij} \ln ( \frac{p_{ij}}{p_{ij}^*} ). \label{relative_entropy}
\end{equation}

Note that the negative caliber is the relative path entropy $\mathcal{D}(\{p_{ij}\} || \{p_{ij}^*\} )$, also known as the Kullbach-Leibler divergence, so maximizing the caliber is equivalent to minimizing the relative path entropy of the perturbed MSM with respect to the unperturbed MSM.\cite{Hazoglou:2015bs}  Maximizing the caliber is in accord with the principle of \textit{minimum relative entropy}, which holds that upon learning new information about a system (for example, that the true equilibrium populations are $\pi_i$, superceding our prior knowledge $\pi_i^*$) the new distribution should be as difficult to discriminate from the original as possible.  In the absence of no prior information (i.e. uniform  $\{p_{ij}^*\}$), minimizing the relative entropy is equivalent to maximizing the absolute path entropy (without a reference).

To maximize the caliber, we turn to a Lagrange multiplier approach to minimize $\mathcal{D}$ with constraints on probability conservation,
\begin{equation}
\sum_j p_{ij} - 1 = 0 \label{probability_constraint}
\end{equation}
for all states $i$, and constraints on detailed balance,
\begin{equation}
\pi_i p_{ij} - \pi_j p_{ji} = 0 \label{detailed_balance_constraint}
\end{equation}
for all pairs of states $i$ and $j$.

With these constraints we can now find new rates $p_{ij}$ that will minimize the relative entropy through a Lagrange multiplier approach.  The Lagrange function to be minimized is

\begin{equation}
\mathcal{L} = \mathcal{D} + \sum_i v_i (\sum_j p_{ij} - 1) + \sum_{i,j} \lambda_{ij}(\pi_i p_{ij} - \pi_j p_{ji}),
\end{equation}
where $v_i$ and $\lambda_{ij}$ are Lagrange multipliers.  First, we solve for optimal $p_{ij}$ in terms of the unknown Lagrange multipliers by setting $\frac{\partial \mathcal{L}}{\partial p_{ij}} = 0 $.

\begin{equation}
\begin{split}
& \frac{\partial}{\partial p_{ij}}  \sum_{i,j} \pi_i p_{ij} \ln ( \frac{p_{ij}}{p_{ij}^*} ) + v_i \frac{\partial}{\partial p_{ij}} (\sum_j p_{ij} - 1)  \\
& + \frac{\partial}{\partial p_{ij}} \sum_{i,j} \lambda_{ij}  (\pi_i p_{ij} - \pi_j p_{ji}) = 0
\end{split}
\end{equation}
which simplifies to 
\begin{equation}
\pi_i \ln (\frac{p_{ij}}{p_{ij}^*}) + \pi_i + v_i + \pi_i(\lambda_{ij}-\lambda_{ji})= 0
\end{equation}
from which we obtain	
\begin{equation}
p_{ij} = p_{ij}^* e^{-\Delta_{ij}} w_i
\end{equation}
Here, for convenience, we define $\Delta_{ij} = \lambda_{ij}-\lambda_{ji}$ and $w_i = e^{-v_i/\pi_i}$.   To determine the value of Lagrange multipliers $\lambda_{ij}$ and $v_i$, we insert our expression $p_{ij}$ into the restraint equations \eqref{probability_constraint} and \eqref{detailed_balance_constraint}.

Using the constraints $\pi_i p_{ij} = \pi_j p_{ji}$ we find that

\begin{equation}
\pi_i p_{ij}^* e^{-\Delta_{ij}} w_i = \pi_j p_{ji}^* e^{+\Delta_{ij}} w_j,
\end{equation}
from which we obtain 
\begin{equation}
e^{-\Delta_{ij}} = \sqrt{  \frac{\pi_j p_{ji}^* w_j}{\pi_i p_{ij}^* w_i} }.
\end{equation}

From the constraint $\sum_j p_{ij} = 1$, we find that 

\begin{equation}
w_i = \frac{1}{\sum_j p_{ij}^* e^{-\Delta_{ij}} }.
\end{equation}

These expressions do not result in a closed-form solution, but rather a simple iterative scheme that numerically converges.  The iterative procedure begins with an initial estimate of $w_i$ (we find that $w_i = 1$ works well in all practical cases we have tried).  We next iteratively estimate
\begin{equation}
p_{ij} \leftarrow p_{ij}^* \sqrt{  \frac{\pi_j p_{ji}^* w_j}{\pi_i p_{ij}^* w_i} } w_i
\end{equation}
and then
\begin{equation}
w_i \leftarrow \frac{w_i}{\sum_j p_{ij}},
\end{equation}
repeating until converged within desired tolerance ($10^{-15}$ in all results we show below).

\section{Results}

\subsection{Biased diffusion models}

\begin{figure}[ht!]
    \includegraphics[width=\columnwidth]{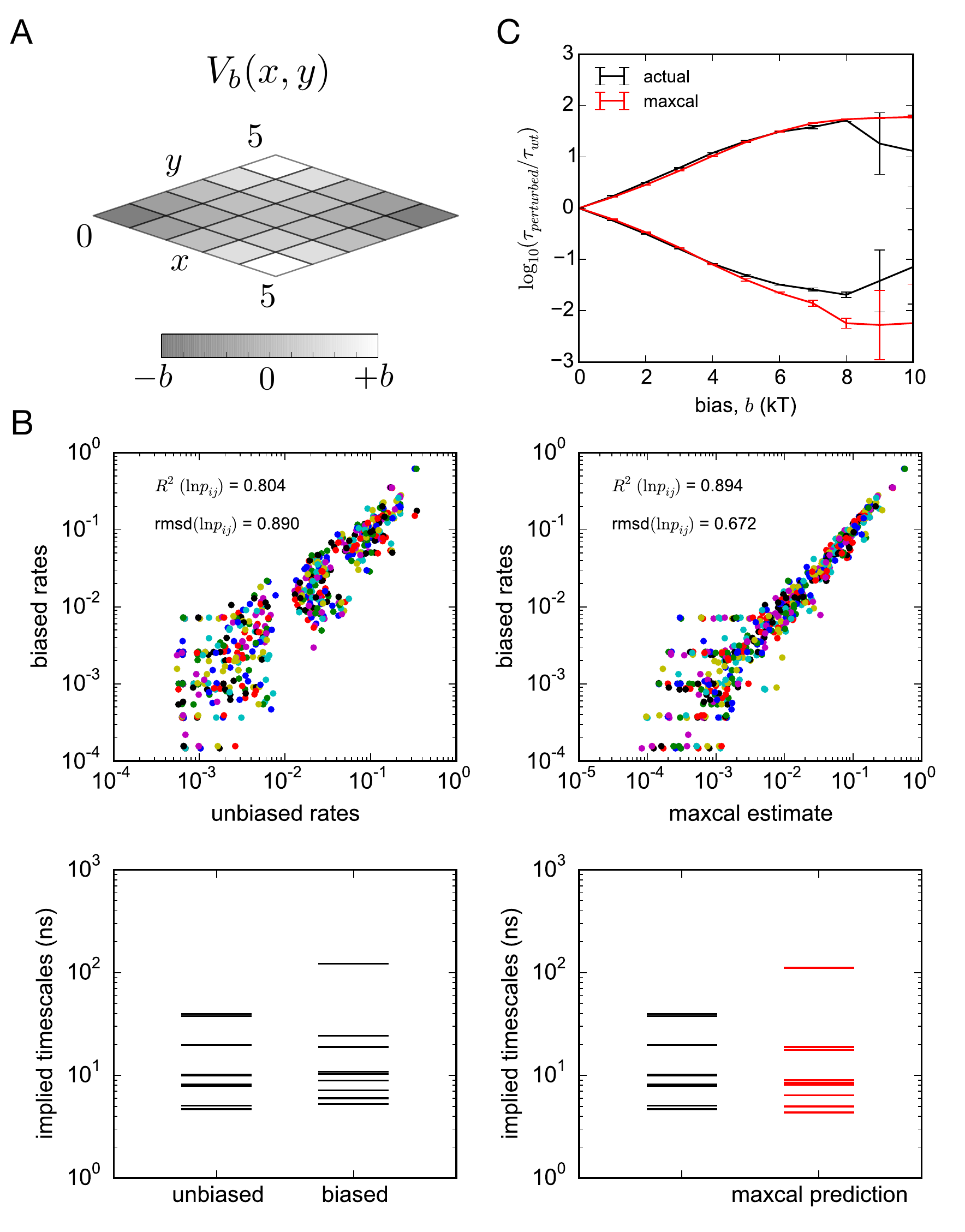}
    \caption{A maximum-caliber method predicts changes in relaxation times for biased 2D diffusion. (A) Biasing potential applied to a 2D diffusion MSM with 25 states, where $b$ is the bias in units $kT$. (B) Left: a comparison of unbiased and biased rates between MSM states, for $b = 2 kT$, shown with MSM implied timescale spectra. Right: a comparison of biased rates with maximum-caliber predictions, shown with implied timescale spectra.  (C) Maximum-caliber predictions of the slowest relaxation timescale are accurate over a range of applied biases, up to two orders of magnitude in rate differences.  In this case, finite sampling ($10^6$ steps) ultimately limits the accuracy of the maximum-caliber predictions beyond biases of $\pm 6$ $kT$.  Error bars are computed from the standard deviation across five independent trials.}
    \label{fig:2D_diffusion}
\end{figure}

We first explore the performance of the maximum-caliber approach in simple biased two-dimensional diffusion models.  As a test system, we consider diffusive dynamics of a particle over $x,y \in (0,5)$ in the presence of a bias potential

\begin{equation}
V_b(x,y) = -\frac{b}{4} (\lfloor x \rfloor - 2)(\lfloor y \rfloor - 2)
\end{equation}

where $b$ is the bias in units $kT$ (Figure \ref{fig:2D_diffusion}a). To construct a Markov State Model (MSM) of the dynamics, a Monte Carlo algorithm was first used to generate a single long dynamic trajectory of particle diffusion.  Proposed moves in $x$ and $y$ were normally-distributed translations taken from $N(\mu=0, \sigma=0.2)$, accepted according to the Metropolis criterion.  MSMs were constructed using a lag time of $\tau_{\text{lag}} = 25$ steps, for which the dynamics is Markovian (Figure \ref{fig:2D_diffusion}b). Discrete MSM states were defined as the set of twenty-five $1 \times 1$ squares tiling the $x,y$-domain.  The matrix of transition probabilities $p_{ij}$ was computed using a maximum-likelihood estimator given the observed transition counts.\cite{Beauchamp:2011he,Wu:2014jy}

Given ``wild type" unbiased diffusion MSMs built from trajectories of $10^6$ steps, we examined the ability of our maximum-caliber approach to predict perturbed transition probabilities and implied timescales for biased diffusion (i.e. the ``mutant").  The implied timescales of an MSM are $\tau_n = -\tau_{\text{lag}}/\ln \mu_n$, where $\mu_n$ are the eigenvalues of the transition probability matrix.  The unbiased MSMs yielded transition probabilities $p_{ij}^*$ and equilibrium populations $\pi_i^*$, from which the biased populations were computed as

\begin{equation}
\pi_i =  \frac{\pi_i^* \exp( -V_i)}{\sum_i \pi_i^* \exp( -V_i)}
\end{equation}

As a detailed example, we present the maximum-caliber results for a bias of $b=2$ kT, shown in Figure \ref{fig:2D_diffusion}b. In this case, the maximum-caliber method predicts estimated transition probabilities that agree more closely with the estimates from the biased MSM (a correlation coefficient of $R^2 = 0.894$, and a root-mean-squared  of 0.672 for predicted $\ln p_{ij}$ values), than unbiased estimates ($R^2 = 0.804$, rmsd($\ln p_{ij}$) =  0.890).  Moreover, the spectrum of implied timescales predicted by maximum-caliber agrees very well with the spectrum from the biased MSM.  Maximum-caliber predictions for a range of bias potentials $0 < b < 10$ kT, in both the forward and backward directions (i.e. predictions of unbiased transition rates from the biased MSM) agree well with the actual MSMs, up to about two orders of magnitude in perturbed rates before succumbing to finite sampling error with trajectories of $10^6$ steps.

\subsection{The maximum-caliber approach predicts changes in mini-protein folding rates due to mutations}

To test the ability of our maximum-caliber approach to predict changes in folding kinetics due to mutations, we utilize MSMs built in previous studies for tryptophan variants of GB1 hairpin,\cite{Razavi:2015bu} and salt-bridge mutations of the Fs peptide helix.\cite{Zhou:2016gb}  A key finding in both these studies was the importance of non-native interactions in shaping folding rates and mechanisms.  For example, according to a simple two-state model of folding, the greater stability of the trpzip4 hairpin compared to the GB1 wild type\cite{Cochran:2001vg} should confer a faster folding rate.  Instead, both experiment \cite{Du:2004hs} and simulation models\cite{Razavi:2015bu} show trpzip4 folds at a slower rate, due to non-native interactions in the unfolded state. Thus, prediction of perturbed folding kinetics due to mutations should be highly non-trivial and provide a stringent test of the maximum-caliber approach.

\begin{figure}[ht!]
    \includegraphics[width=\columnwidth]{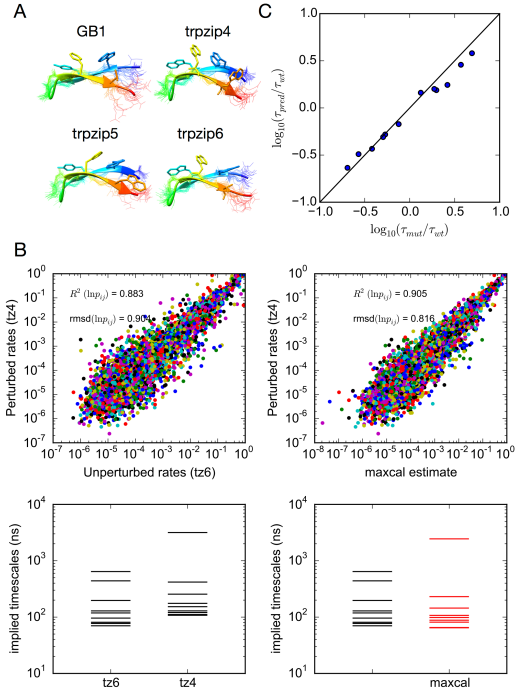}
    \caption{A maximum-caliber method predicts changes folding kinetics for tryptophan variants of GB1 hairpin  (A) Visualization of conformations taken from the native macrostate of MSMs constructed for GB1 hairpin variants, as described in Razavi and Voelz.\cite{Razavi:2015bu}.  (B) Left: a comparison of rates between MSM states for trpzip6 (unperturbed) and trpzip4 (perturbed), shown with MSM implied timescale spectra. Right: a comparison of the trpzip4 rates with maximum-caliber predictions from perturbing the trpzip6 MSM, shown with implied timescale spectra.  (C) Comparisons of the actual and predicted differences in the folding relaxation timescales, across all twelve possible sequence comparison, agree over a range of more than an order of magnitude.}
    \label{fig:hairpin}
\end{figure}

\paragraph*{GB1 hairpin and tryptophan variants.}
Four 150-macrostate MSMs for GB1, trpzip4, trpzip5 and trpzip6 (Table \ref{table:hairpin_sequences} and Figure \ref{fig:hairpin}a) were 
constructed from over 9 ms of explicit-solvent trajectories simulated on the Folding@home distributed computing platform, according to the methods described in Razavi and Voelz.\cite{Razavi:2015bu}  These MSMs were carefully constructed from the combined trajectory data of all four sequences, using tICA-based clustering and subsequent macrostate coarse-graining, to achieve a metastable state decomposition that could accurately represent all four hairpin designs.  To test our maximum caliber approach, we used the equilibrium populations $\pi_i^*$ and transition probabilities $p_{ij}^*$ from a wild type MSM to infer perturbed transition probabilities and folding kinetics of a mutant MSM with equilibrium populations $\pi_i$.   The four sequences (gb1, trpzip4, trpzip5, trpzip6) provide twelve different pairwise predictions to the test the maximum-caliber method.

As a specific example, we show results for predicting the folding kinetics of trpzip4 from the MSM of trpzip6 (Figure \ref{fig:hairpin}b). Our previous simulation results showed that this V14W mutation induces a dramatic increase in the folding time, by almost an order of magnitude.  Maximum-caliber predicts the same result.  As in the 2D diffusion example above, maximum-caliber estimates of the transition probabiilties are more correlated with the trpzip4 MSM  ($R^2 = 0.905$, rmsd($\ln p_{ij}$) =  0.816) than with the wild type MSM  ($R^2 = 0.883$, rmsd($\ln p_{ij}$) =  0.904), although in both cases there is increased variability due to the complexity of the models (150 macrostates).  Implied timescale predictions agree very well with actual MSM timescales.

\begin{table}[h]
\resizebox{\columnwidth}{!}{\begin{minipage}{\columnwidth}
\begin{tabular}{lcr}
abbreviation              & sequence \\
\hline
GB1 &  \texttt{GEWTYDDATKTFTVTE} \\
trpzip4 &  \texttt{GEWTWDDATKTWTWTE} \\	
trpzip5 &  \texttt{GEWTYDDATKTFTWTE} \\
trpzip6 &  \texttt{GEWTWDDATKTWTVTE} \\

\end{tabular}
\caption[Table caption text]{Sequences of GB1 hairpin variants.}
\label{table:hairpin_sequences}
\end{minipage}}
\end{table}

\begin{figure}[ht!]
    \includegraphics[width=0.6\columnwidth]{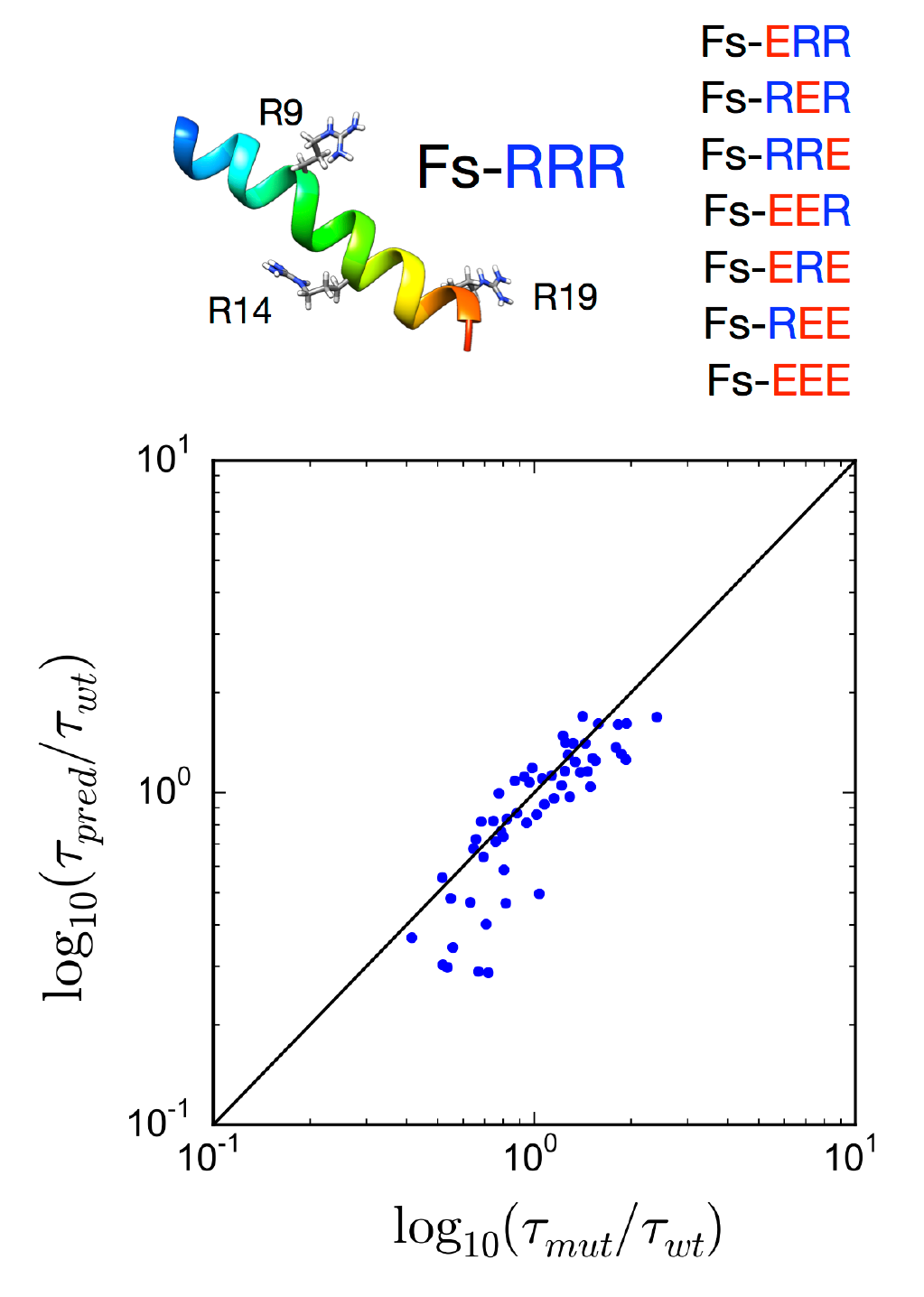}
    \caption{The maximum-caliber method predicts changes in folding kinetics for salt-bridge mutations of Fs-peptide helix.  Comparisons of the actual and predicted differences in the folding relaxation timescales, across all 54 possible sequence comparisons, correlate well over a range of about an order of magnitude. }
    \label{fig:Fs-peptide}
\end{figure}

\paragraph*{Fs peptide helix and salt-bridge mutants.}
MSMs for salt-bridge mutants of Fs peptide (Ace-A$_5$AAARAAAARAAAARAA-Nme) provide another valuable test of the maximum-caliber approach (Figure \ref{fig:Fs-peptide}a).  As described in Zhou and Voelz,\cite{Zhou:2016gb} eight Fs peptide helix variants, in which the three arginine residues were mutated in all possible combinations with glutamic acid, were simulated on Folding@home in explicit solvent to produce about 130 $\mu$s of total trajectory data per sequence.   The glutamic acid mutations were designed to introduce potential salt-bridge interactions that can only be formed in unfolded states, inducing highly non-trivial changes in folding kinetics from non-native interactions.   

Like the hairpin systems above, the Fs peptide MSMs were constructed using tICA-based clustering of the combined trajectories of all sequences, producing a 1200-microstate MSM with  lag time of 5 ns.  Optimal MSM construction parameters such as the tICA lag time, number of tICA components, number of microstates and clustering method were chosen systematically using GMRQ variational cross-validation.\cite{McGibbon:2015gm} In our original study, microstate MSMs were lumped into a combined 40-macrostate models using the BACE algorithm\cite{Bowman:2012jz}; here, we present 30-macrostate MSMs lumped using the same method that retain nearly identical kinetics.   The additional lumping helps improve the overlap in metastable states for the eight sequences. 

The eight Fs peptides sequences provide a total of 54 different wild type vs. mutant predictions to  test the maximum-caliber method.  MSM folding times versus maximum-caliber predictions show good agreement across this large number of comparisons (Figure \ref{fig:Fs-peptide}).

\subsection{The maximum-caliber approach predicts changes in folding kinetics for a WW domain mutation.}

WW domain is a small three-stranded beta-sheet protein whose folding kinetics has been studied extensively by experiment\cite{Nguyen:2003gw,Nguyen:2005ec,Jager:2006ga,Liu:2008fs} and molecular simulation.\cite{Freddolino:2008dj,Noe:2009en,Shaw:2010ge,LindorffLarsen:2011gl}. Protein engineering studies have discovered many fast-folding variants of WW domain, most notably the FiP35 sequence, which has a folding time of 13 $\mu s$.\cite{Nguyen:2005ec} An even faster-folding (3 $\mu s$) variant of WW domain was subsequently discovered by Piana et al. through molecular simulation-based computational prediction.\cite{Piana:2011gm}   In the fast-folding variant, the native sequence Asn-Ala-Ser (NAS) near loop 2 is replaced with Gly-Thr-Thr (GTT), a sequence whose backbone propensities are more favorable for the native conformation.  The discovery of the GTT variant is significant because it was predicted using direct simulation of reversible folding on the Anton supercomputer,\cite{Shaw:2008go} the first such example of this approach for protein design.\cite{Piana:2011gm}   Subsequent temperature-jump refolding experiments showed stabilities and folding rates in good agreement with the predictions, validating the computational design.

We wanted to see if our maximum caliber approach could be used to make a similar prediction, without the need to perform expensive brute-force folding simulations.   Towards this end, we built Markov State Models of the FiP35 and GTT variants of WW domain using published trajectory data from Shaw et al., to see if our maximum caliber approach could correctly predict changes folding rates given estimates of the differences in equilibrium probabilities.

\begin{figure}[ht!]
    \includegraphics[width=0.75\columnwidth]{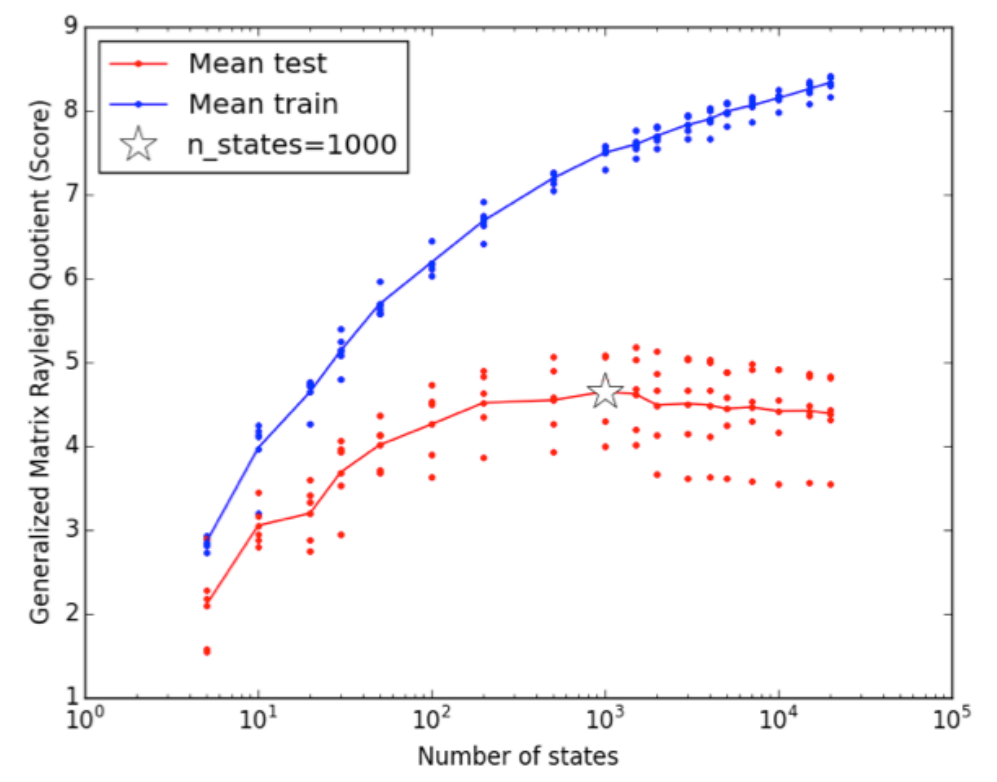}
    \caption{GMRQ results for the GTT WW domain 1000-microstate ``solo'' MSM.}
    \label{fig:SI-GMRQ}
\end{figure}

\begin{figure}[ht!]
    \includegraphics[width=0.75\columnwidth]{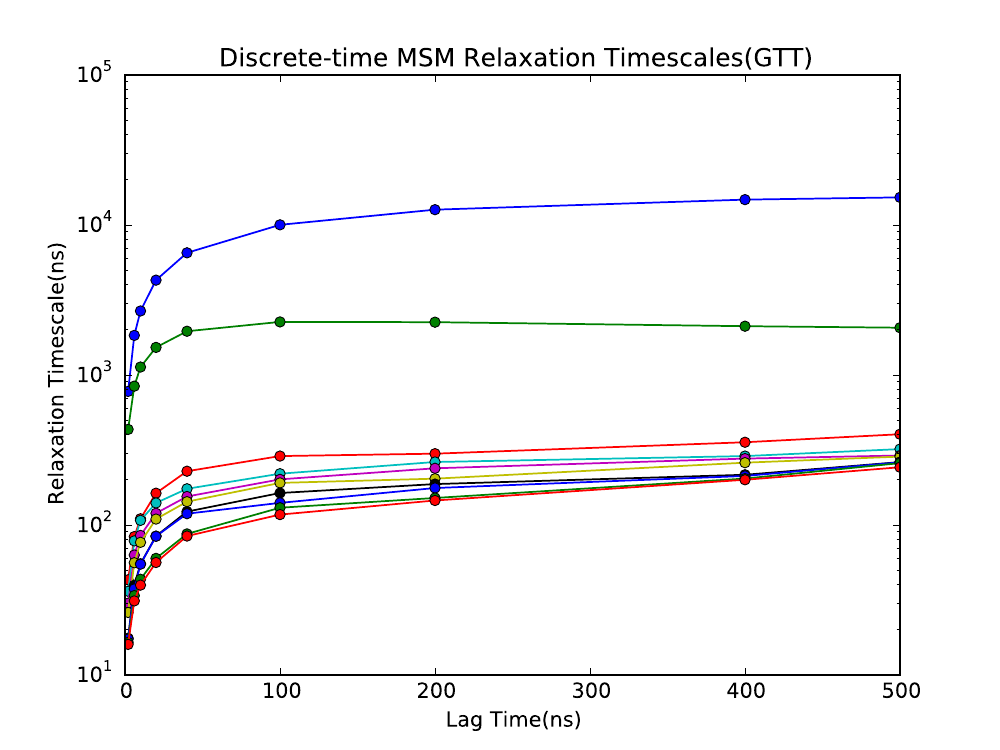}
    \caption{Implied timescales versus lag time for the 1000-microstate ``combo" MSM of GTT WW domain.}
    \label{fig:SI-implied-timescales-GTT}
\end{figure}

\begin{figure}[ht!]
    \includegraphics[width=0.75\columnwidth]{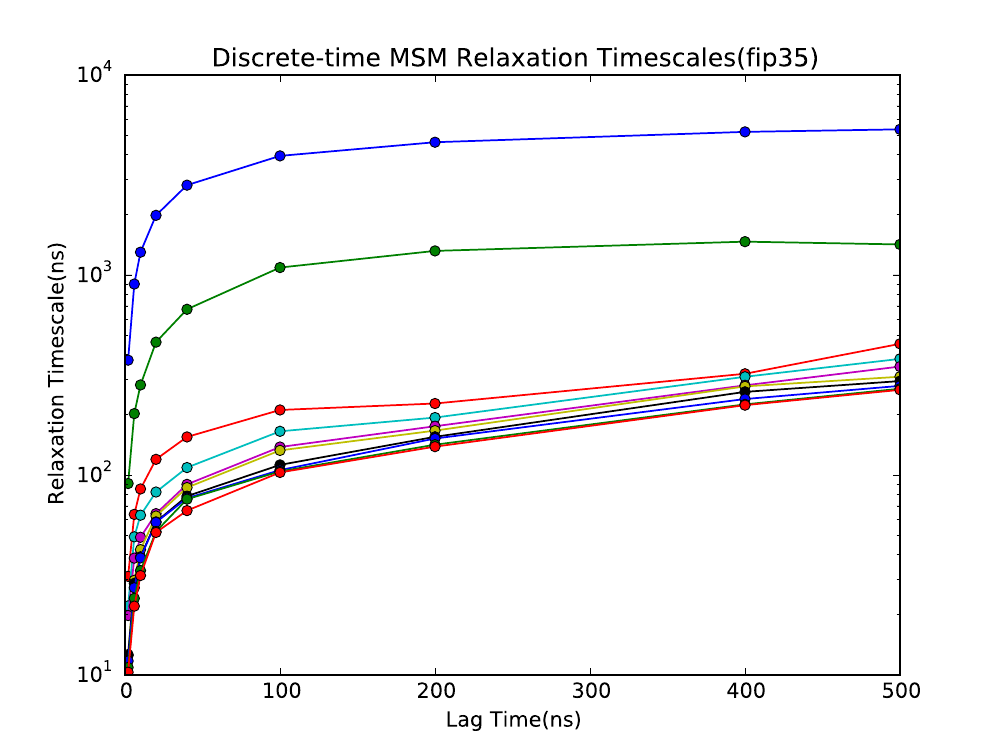}
    \caption{Implied timescales versus lag time for the 1000-microstate ``combo" MSM of FiP35 WW domain.}
    \label{fig:SI-implied-timescales-FiP35}
\end{figure}

\paragraph*{Construction of MSMs for WW domain folding.}
Trajectory data for GTT WW domain came from two independent simulation trajectories of lengths 651 $\mu s$ and 486 $\mu s$, performed at 360 K.\cite{LindorffLarsen:2011gl}  Trajectory data for FiP35 WW domain came from two $\sim$100 $\mu$s trajectories performed at 395 K.\cite{Shaw:2010ge}.    In all MSMs we describe below, we used tICA with a lag time of 10 ns to find the best low-dimensional subspace projection to perform $k$-centers clustering. All pairwise distances between C$_\alpha$ and C$_\beta$ atoms were used as the input coordinates for tICA.   The GMRQ method\cite{McGibbon:2015gm} was used to determine that 1000 microstates and 8 tICA components is optimal for accurately capturing folding dynamics (Figure \ref{fig:SI-GMRQ}). An MSM lag time of 100 ns was chosen for all models, based on the observation that computed implied timescales plateau near this lag time (Figures \ref{fig:SI-implied-timescales-GTT} and \ref{fig:SI-implied-timescales-FiP35}).

To test the assumption that the relevant metastable states are conserved for both GTT and FiP35, several methods were used to build MSM models.  First, we built individual MSMs for each sequence, performing a separate tICA analysis and microstate clustering for each.  We will refer to these as the ``solo" models of GTT and FiP35.  Second, we performed a tICA analysis on the combined data of GTT and FiP35, and then built separate MSMs for GTT and FiP35 using this tICA projection.  We will refer to these as the ``projected" models of GTT and FiP35. Finally, we built MSMs from the combined data of GTT and FiP35, using the combined tICA projection to performing microstate clustering on the combined data.  Separate MSMs for GTT and FiP35 were then built using these combined microstate generators, such that both shared metastable state definitions.  We refer to these models as the "combined" MSMs for GTT and FiP35; because they share metastable states, these models are amenable to our maximum-caliber approach.  

Due to finite sampling, however, not all metastable states are sampled by each sequence.  Since states with zero population presents problems for the maximum caliber estimator, we perform further coarse-graining of our 1000-microstate MSM to a 250-macrostate MSM using the BACE algorithm.\cite{Bowman:2012jz}  Using this procedure we obtain a set a metastable states that are sampled by each sequence, although some states are visited so rarely as to cause long timescale artifacts.  These states $j$ can easily be identified by estimated transition probabilities $T_{ij} < 10^{-20}$, which we remove by setting to zero and perform ergodic trimming using Tarjan's algortihm.\cite{Scalco:2011dz}  We refer to the 250-macrostate MSMs as the ``lumped" models.

\begin{figure}[ht!]
    \includegraphics[width=\columnwidth]{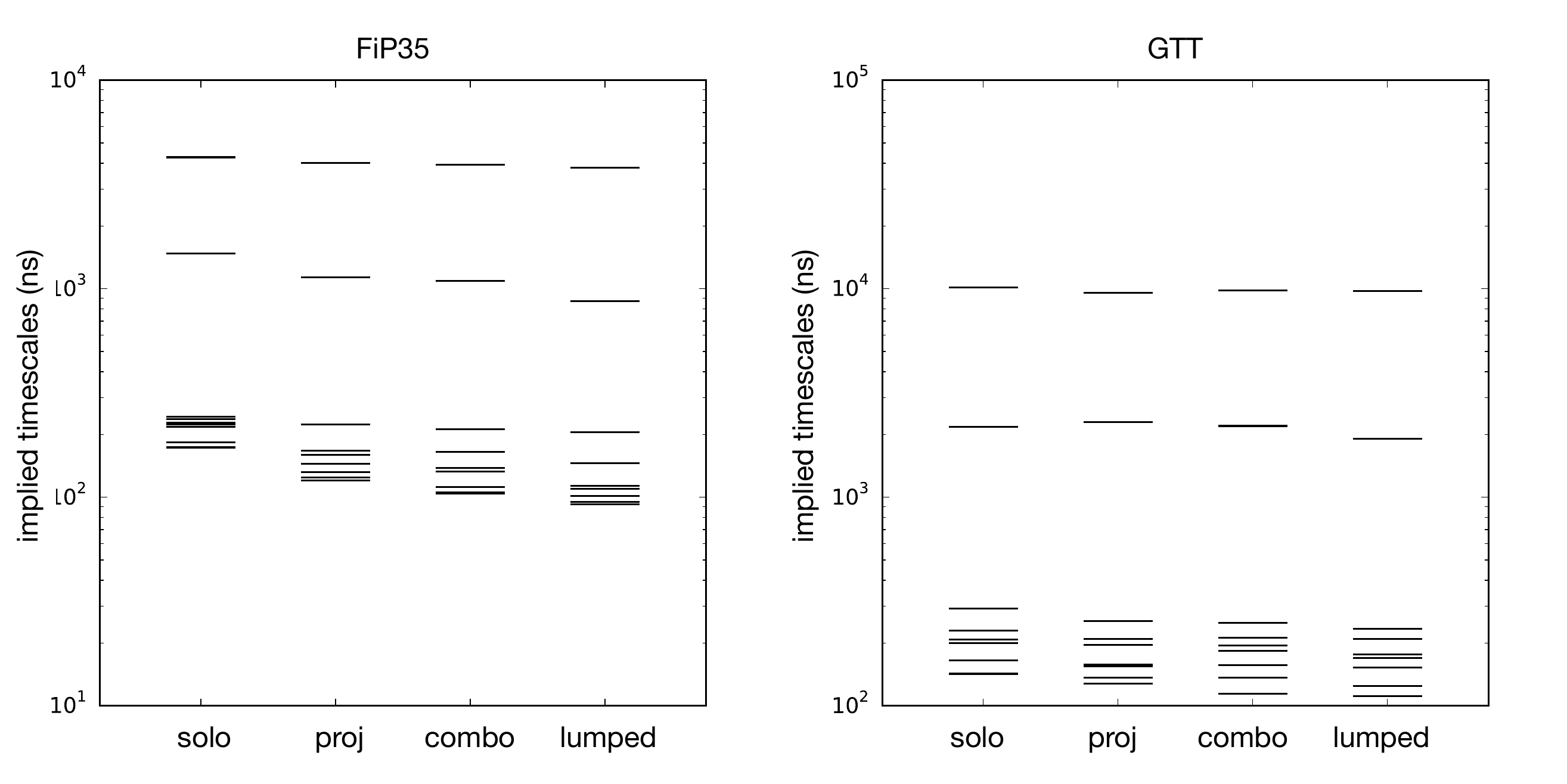}
    \caption{Implied timescales of ``solo",``projected", ``combo" and ``lumped" MSMs for FiP35 and GTT WW domains show minimal losses in quality from projection and coarse-graining.}
    \label{fig:SI-solo_proj_combo_implied_timescales}
\end{figure}

\paragraph*{MSM predictions of WW domain folding timescales.}
The folding dynamics predicted by all of the constructed MSMs recapitulate previously published results.  The solo, projected, and combined MSMs of FiP35 WW domain predict folding relaxation timescales of 4.3, 4.0, 3.9 and 3.8, respectively, which each successive model suffering only slightly from projection artifacts (Figure \ref{fig:SI-solo_proj_combo_implied_timescales}) .  To put these predicted timescale into perspective, it is useful to note that Shaw et al. reported folding times of 10 $\pm$ 3 $\mu$s, based on the average waiting time in the unfolded state seen in the trajectory data (at 395 K),\cite{Shaw:2010ge} in comparison to the experimental folding time of 14 $\mu$s measured at 340 K.\cite{Liu:2008kj}  Previous MSMs built by Beauchamp et al. from this data using a lag time of 50 $\mu$s yield a predicted folding relaxation time of $\sim$2 $\mu$s.\cite{Beauchamp:2012kp}   Beauchamp et al. also used a self-consistent rate estimator (SCRE) derive a folding relaxation time of $\sim$9 $\mu$s for a three-state model of Fip35 folding.  This range was consistent with estimates of $\sim$5 $\mu$s (using a 100 ns lagtime) from Lane et al.,\cite{Lane:2011fp} and estimates of $\sim$3.5 $\mu$s (using a 75 ns lag time)  from McGibbon and Pande,\cite{McGibbon:2013bz} which tested improved MSM construction methods.

The solo, projected, combined and lumped MSMs of GTT WW domain predict folding relaxation timescales of 10.2, 9.6, 9.8  and 9.8, respectively.  Shaw et al. reported a folding time of 21 $\pm$ 6 $\mu$s from the average unfolded state lifetime seen in the trajectory data (at 360 K), in comparison to the experimental folding time of 5.7 $\mu$s.\cite{Piana:2011gm}.   Previous MSMs built by Beauchamp et al. from this data using a lag time of 50 $\mu$s yield a predicted folding relaxation time of $\sim$6 $\mu$s.\cite{Beauchamp:2012kp}, and a SCRE estimate of $\sim$8 $\mu$s for a three-state model of GTT folding. Our predictions are larger than these values, likely reflecting improvements in MSM construction protocols since that work.  

Importantly, we note that all of our MSMs, as do the Beauchamp et al. MSMs, predict FiP35 to have a \textit{faster} predicted folding rate than GTT, which is the opposite result obtained from experiments and the simulation predictions presented in Piana et al.\cite{Piana:2011gm} The explanation for this is that the Piana et al. predictions incorporated additional simulation data unavailable to Beauchamp et al. at the time, namely: four additional simulations of GTT (trajectory lengths of 83 $\mu$s, 118 $\mu$s, 124 $\mu$s, and 272 $\mu$s) and four additional 100-$\mu$s simulations of FiP35.  Our analysis does not include this additional data either.  We comment here that the data-dependent variability of folding rates calculated from small numbers of trajectories does not bode well for the prospect of using single simulation trajectories to accurately predict folding rates.  In the case of WW domain, due to the long dwell times in folded and unfolded states, the correct result was obtained only after an additional millisecond of molecular simulation could be performed. This underscores the advantages of MSMs for studying conformational dynamics.

\begin{figure}[ht!]
    \includegraphics[width=\columnwidth]{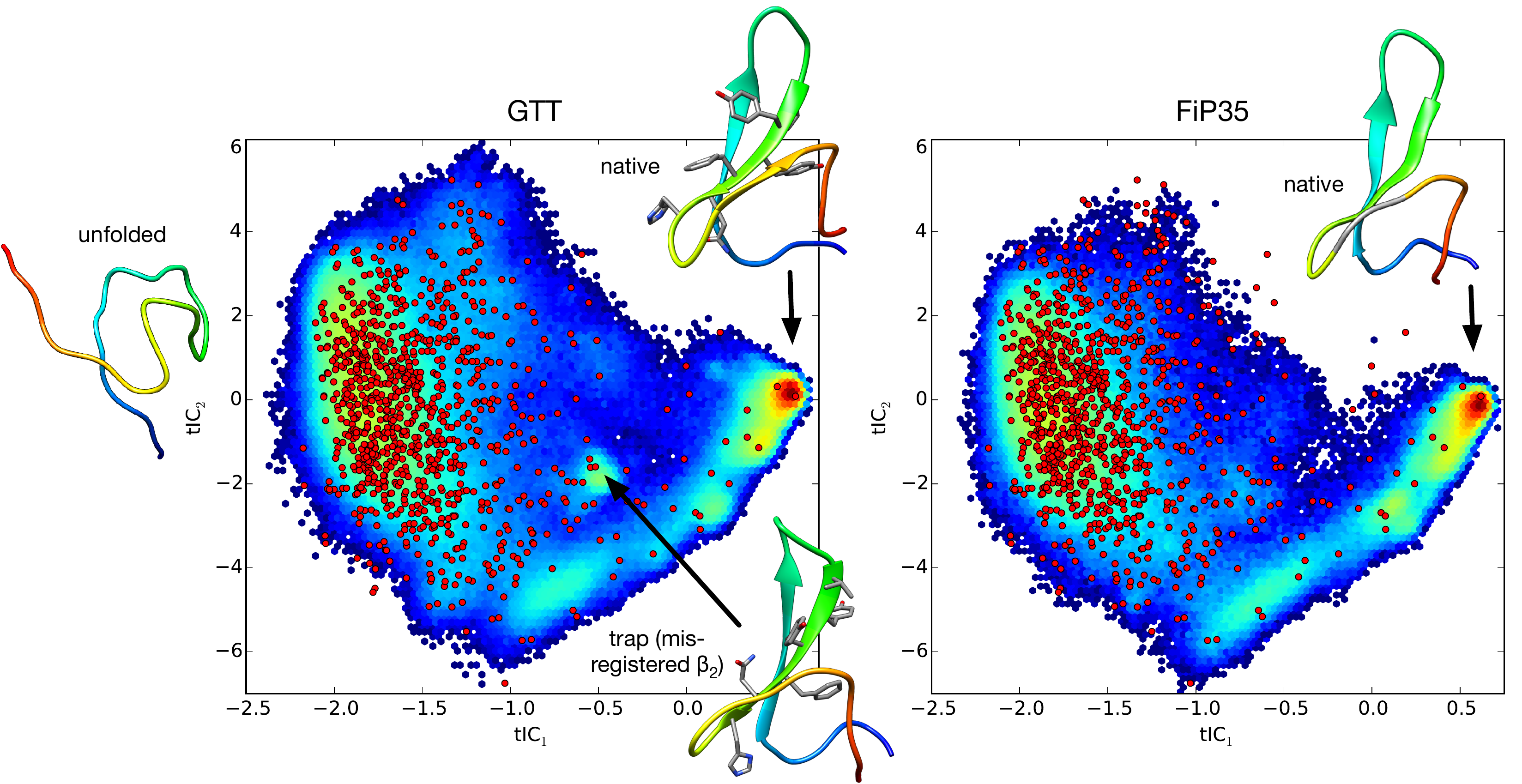}
    \caption{Projection of GTT and FiP35 trajectory data to the two largest tICA components, overlaid with the (shared) locations of the 1000 ``combo'' MSM microstates (red dots).  The two folding landscapes are highly similar, although unlike FiP35, GTT shows significant population for a non-native trap conformation.}.
    \label{fig:WW-tica}
\end{figure}

\begin{figure}[ht!]
    \includegraphics[width=\columnwidth]{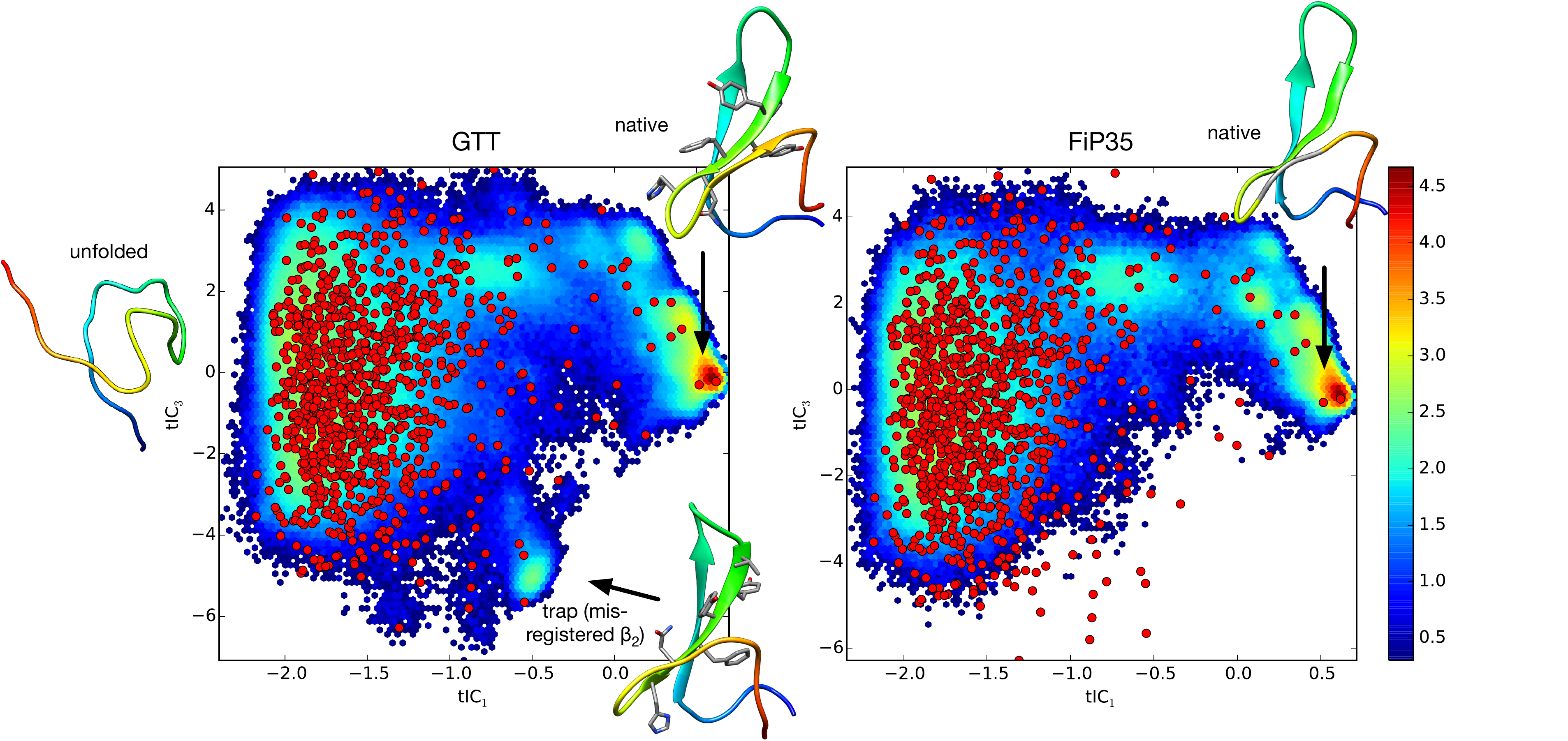}
    \caption{Projection of GTT and FiP35 trajectory data to tIC$_1$ and tIC$_3$ components, overlaid with the (shared) locations of the 1000 ``combo'' MSM microstates (red dots).  The tIC$_3$ reveals the significance of the non-native trap conformation in determining the second-slowest relaxation timescale. Inspection of of the eigenvector structure for this relaxation (not shown) confirms that the trap is off-pathway to folding.}
    \label{fig:SI-tICA3_landscapes_WWdomain}
\end{figure}

\paragraph*{MSM predictions of WW domain folding mechanism.}
Projection of the 1000-microstate "combined" MSMs onto the two largest tICA components (tIC$_1$ and tIC$_2$) reveal similar folding mechanisms of GTT and FiP35 variants.  The tIC$_1$ component corresponds to the slowest (folding) relaxation timescale; along this component a broad unfolded-state basin is separated from narrow folded-state basin (Figure \ref{fig:WW-tica}).  The next-slowest relaxation corresponds to an off-pathway "trap" state, as further revealed by projections to tIC$_1$ and tIC$_3$ (Figure \ref{fig:SI-tICA3_landscapes_WWdomain}). This trap state has a native-like fold, but with the second beta strand inverted with mis-registered hydrogen bonds.  In this sense, the trap state is a near-native "decoy" on the free energy landscape, i.e. a local, not global, free energy minimum.  Similar to the reports of Beachamp et al,\cite{Beauchamp:2012kp} we see a high population of this trap state for the GTT variant, but not the FiP35 variant.  This may be due to particular sequence-dependent effects, differences in force fields (FiP35 trajectories used the AMBER ff99SB-ildn force field, while GTT trajectories used CHARMM22*), or differences in conformational sampling.  The total simulation time for the GTT trajectories is over five times that of the FiP35 data, which is evident from the larger sampled area of the tICA projection for GTT.

\begin{figure}[ht!]
    \includegraphics[width=\columnwidth]{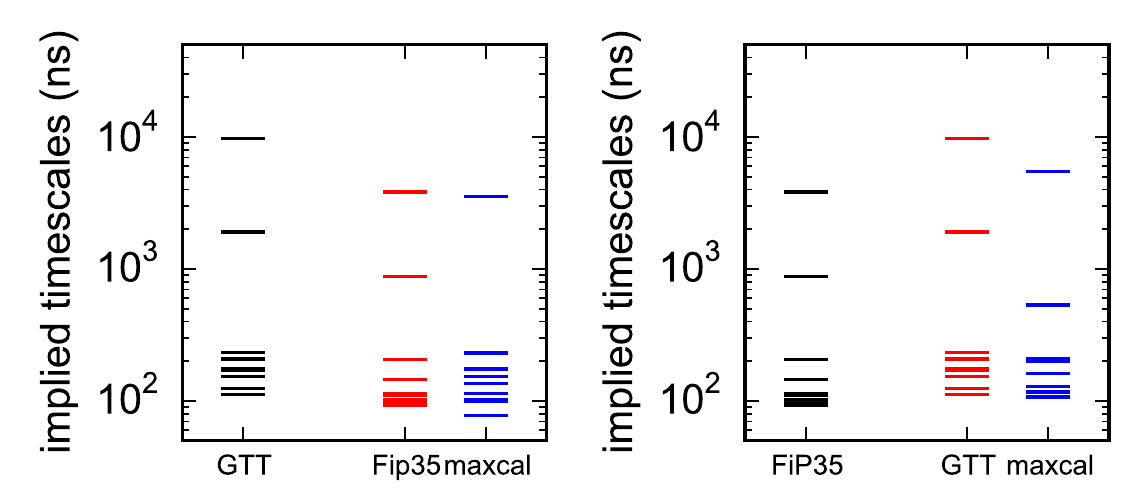}
    \caption{Maximum-caliber predictions of FiP35 folding timescales from the  250-macrostate MSM of GTT (left), and GTT folding timescales from the  250-macrostate MSM of FiP35 (right), agree well with the actual values. The latter results are less accurate, which may be attributed to the smaller amount of trajectory data used to construct the FiP35 model ($\sim$5x less compared to GTT).}
    \label{fig:WW-maxcal}
\end{figure}

\paragraph*{Maximum-caliber predictions of folding rates.}
We applied our maximum-caliber approach to predict the folding rates of GTT WW domain given the 250-macrostate "lumped" MSM populations of Fip35 WW domain, and vice versa.  The results show that this approach successfully predicts the extent to which the Gly-Thr-Thr versus Asn-Ala-Ser mutations perturb the slowest folding relaxation (Figure \ref{fig:WW-maxcal}).  The maximum-caliber predictions of the FiP35 folding relaxation timescale match the true values very closely.  The computed GTT folding relaxation timescale is less accurate, although it correctly predicts that the GTT folding relaxation timescale is slower than FiP35. These differences in accuracy between the FiP35 and GTT predictions might be attributed to the greater amount of available GTT trajectory data.

\begin{figure}[ht!]
    \includegraphics[width=0.8\columnwidth]{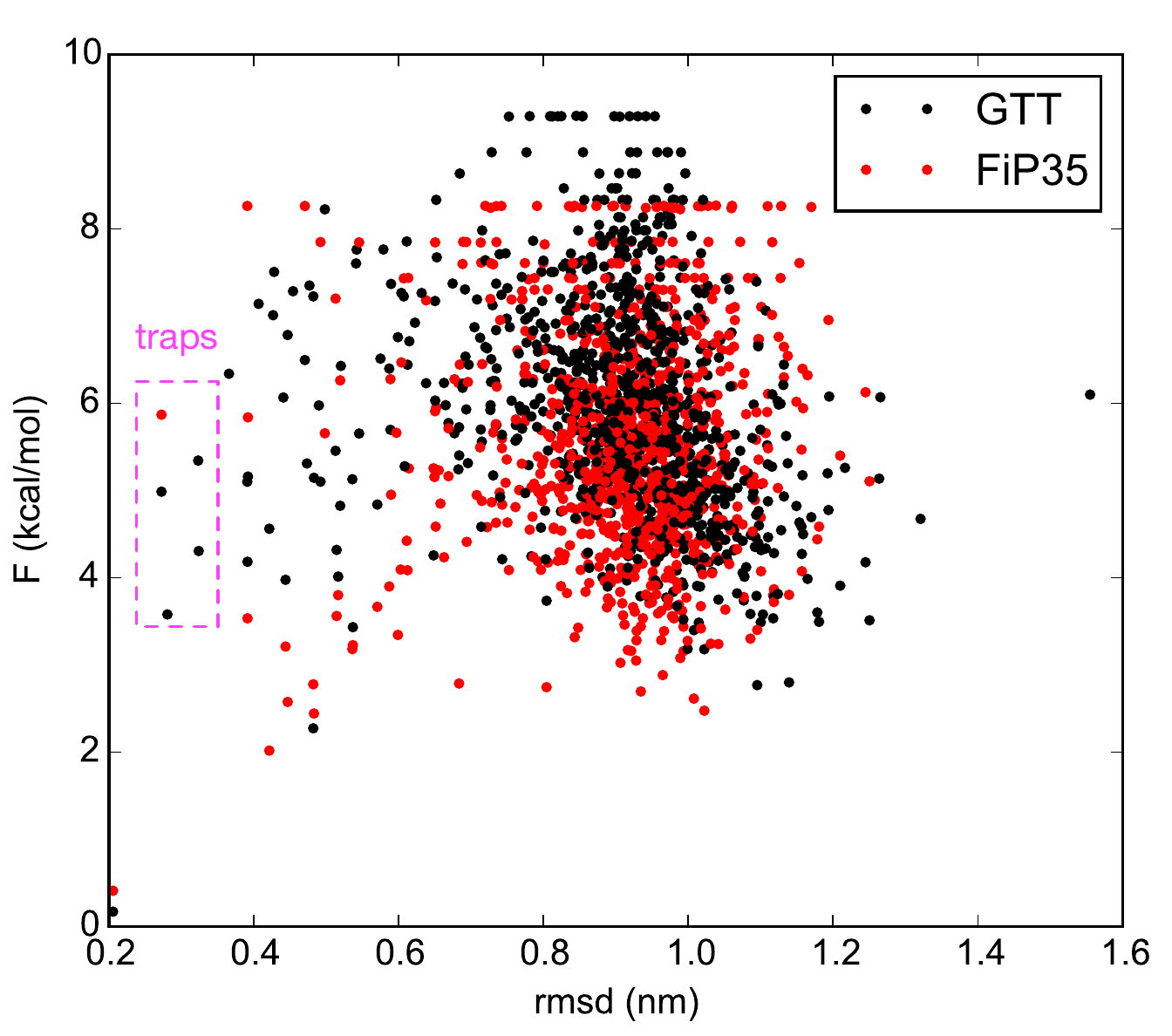}
    \caption{Microstate free energies $F_i = -kT \ln \pi_i$ from the 1000-microstate ``combo'' MSMs show that the unfolded state and on-pathway transition states of FiP35 are stabilized with respect to GTT, resulting in the accelerated folding rate.  Off-pathway trap states are destabilized for FiP35 compared to GTT.}
    \label{fig:SI-F_vs_rmsd}
\end{figure}

A comparison of the free energies of microstate populations  $F_i = -kT \ln \pi_i$ for the FiP35 and GTT 1000-microstate ``combo'' MSMs, plotted as a function of the average backbone rmsd (root-mean-squared deviation of N, CA, C, and O atoms) to a randomly chosen native-state microstate conformation reveal that, in general, the FiP35 state populations have stabilized unfolded basins and transition states compared to the native state, resulting in accelerated folding rates (Figure \ref{fig:SI-F_vs_rmsd}). Unlike the 2D biased diffusion and GB1 hairpin variant models, the next-slowest relaxation timescales (involving the trap state) are not well-predicted by maximum-caliber.  We suspect this may also be attributed to sampling deficiencies, as the GTT variant is the only sequence that significantly populates the trap state.

\section{Discussion}

The success of our maximum-caliber approach appears to be dependent on several factors, including (1) a metastable state decomposition that closely reflects the underlying kinetic distances, (2) small enough perturbations as to conserve the important metastable states, and (3) sufficient sampling to make accurate estimates of rate changes. Only recently have such considerations been quantitatively incorporated into standard MSM construction protocols for protein conformational dynamics, which means one can expect wider applicability of this method to more MSMs in the future.

The potential applications of our maximum-caliber method are especially exciting.   For one, with a sufficiently accurate method of predicting mutational free energy changes for each metastable state, the method could be used to elucidate how disease- or resistance-associated mutations perturb protein dynamics.   Similarly, the effects of many candidate mutations could be efficiently interrogated for the purposes of protein design.

More generally, there are many situations where it may be desirable to remove thermodynamic bias from a given MSM, which now can be robustly performed using our maximum-caliber method.  One application would be to construct MSM models of ligand dissociation; unbinding transitions otherwise too rare to observe could be sampled used a biasing potential, and later reweighted by our maximum-caliber approach to calculate unbiased dissociation rates.  Similarly, the common problem of force field bias in molecular simulations could be remedied by using our maximum-caliber approach to reweight MSM transition rates to match experimental observables.

Our maximum-caliber approach is similar in some ways to the dTRAM method,\cite{Wu:2014jy} which uses a Lagrangian method to infer maximum-likelihood estimates of transitions rates from transition counts observed in multiple thermodynamic ensembles.  Perhaps deeper connections exist between these two methods that could be explored in future work, including the incorporation of multiple thermodynamic biases.  Interestingly, the dTRAM algorithm is unable to make estimates of rates for ensembles in which no transition counts are observed.  The maximum-caliber method we present here has no such limitation.

\section{Conclusions}

In this work, we have presented a simple and robust maximum-caliber method for inferring transition rates of a Markov State Model (MSM) with perturbed equilibrium populations, given estimates of state populations and rates for an unperturbed MSM. Applying this approach to several MSMs of protein folding sequence variants results in good predictions of perturbed folding rates directly from changes in equilibrium state populations.

\section*{Acknowledgments}
The authors thank the participants of Folding@home, without whom this work would not be possible.  We graciously acknowledge D.E.Shaw Research for providing access to the WW domain folding trajectory data. This research was supported in part by the National Science Foundation
 through major research instrumentation grant number CNS-09-58854 and MCB-1412508.

\bibliography{relentropy}



\newpage


\newpage
\beginsupplement
\renewcommand{\thepage}{S\arabic{page}}  
\setcounter{page}{1}

\section{Supporting Information} 

\subsection{Supporting Figures}

\end{document}